# RATIONAL DESIGN OF PHOTO-ELECTROCHEMICAL HYBRID DEVICES BASED ON GRAPHENE AND *Chlamydomonas reinhardtii* LIGHT-HARVESTING PROTEINS

Martha Ortiz-Torres[1,2], Miguel Fernández-Niño[1], Juan C Cruz[3], Andrea Capasso[4], Fabio Matteocci[5] Edgar J. Patiño[6], Yenny Hernández[2*], Andrés Fernando González Barrios[1*]

[1]Grupo de Diseño de Productos y Procesos (GDPP), Department of Chemical Engineering, Universidad de los Andes, Bogotá 111711, Colombia.

[2]Nanomaterials Laboratory, Physics Department, Universidad de Los Andes, Bogotá 111711, Colombia.

[3]GINIB Research Group, Department of Biomedical Engineering, Universidad de Los Andes, Bogotá 111711, Colombia.

[4]International Iberian Nanotechnology Laboratory, 4715-330 Braga, Portugal

[5]C.H.O.S.E – Centre for Hybrid and Organic Solar Energy, Department of Electronic Engineering, University of Rome Tor Vergata, Via del politecnico 1, Rome, 00133, Italy

[6]Superconductivity and Nanodevices Laboratory, Physics Department, Universidad de Los Andes, Bogotá 111711, Colombia.

[*]Corresponding authors:

Yenny Hernández (email: yr.hernandez@uniandes.edu.co)

Andrés Fernando González Barrios (email: andgonza@uniandes.edu.co)

**ABSTRACT**

Dye-sensitized solar cells (DSSCs) have been highlighted as the promising alternative to generate clean energy based on low *pay-back time* materials. These devices have been designed to mimic solar energy conversion processes from photosynthetic organisms (the most efficient energy transduction phenomenon observed in nature) with the aid of low-cost materials. Recently, light-harvesting complexes (LHC) have been proposed as potential dyes in DSSCs based on their higher light-absorption efficiencies as compared to synthetic dyes. In this work, photo-electrochemical hybrid devices were rationally designed by adding for the first time Leu and Lys tags to heterologously expressed light-harvesting proteins from *Chlamydomonas reinhardtii*, thus allowing their proper orientation and immobilization on graphene electrodes. The light-harvesting complex 4 from *C. reinhardtii* (LHC4) was initially expressed in *Escherichia coli*, purified via affinity chromatography and subsequently immobilized on plasma-treated thin-film graphene electrodes. A photocurrent density of 40.30 ± 9.26 $\mu A/cm^2$ was measured on devices using liquid electrolytes supplemented with a phosphonated viologen to facilitate charge transfer. Our results suggest that a new family of graphene-based thin-film photovoltaic devices can be manufactured from rationally tagged LHC proteins and opens the possibility to further explore fundamental processes of energy transfer for biological components interfaced with synthetic materials.

**INTRODUCTION**

Decarbonization of the global energy production represents one of the major issues in current research[1], which has resulted in the extensive development of fuels-free technologies[2]. As the highly explored technology, solar cells have been shown to be the most promising solution to address the carbonization problem for which three generations of these devices have been sequentially developed throughout the last half-century[3]. Within the third generation of photovoltaic solar cells, dye-sensitized solar cells (DSSCs) have been proposed as the future devices as a consequence of both a low *pay-back time* and a low *carbon footprint* manufacture. Also, their ability to mimic the photosynthetic transduction of photons into electrons has been considered a key attribute for an exceptional performance[4,5]. Moreover, the capability to separate and reuse their components made DSSCs a remarkable eco-friendly alternative for solar energy utilization.

Recent studies on DSSC performance have shown energy conversion efficiencies of around 14%[6], which could be further improved by fine-tuning the assembly of dyes and electrodes, which are considered critical components for superior transduction[7]. In order to manufacture a DSSC, dyes, and electrodes are required to allow light-induced charge separation through the conjugation interface. Recently, the advantage of using eco-friendly materials as potential dyes and electrodes has been reviewed[8,9]. Among the emerging eco-friendly dyes, photosynthetic proteins have been recently used to design photosynthetic machinery-based DSSCs[9]. In this regard, light-harvesting complexes (LHC) have been proposed as potential dyes for the production of DSSCs due to their ability to efficiently capture solar radiation in the visible region[10–12]. Core reaction centers and whole photosystems have been also incorporated into hybrid solar transduction devices as the main DSSCs dyes [9]. Nevertheless, LHCs, as single proteins, have been shown to maintain their ability for light capturing in isolation[13] i.e., in the absence of interactions within the monomeric subunits. This independence represents an easier/inexpensive alternative to engineer DSSCs.

LHC proteins have been obtained from different photosynthetic organisms including bacteria such as *Rhodobacter spp.* and higher plants such as. *Pisum sativum* and *Spinacia spp.* [9,14–16]. Surprisingly, the utilization of LHCs derived from algae in DSSCs has been poorly explored despite the abundant information available regarding the properties and energy transduction capabilities of the photosystems of these organisms[17–19]. For instance, the photosystems from the model algae *C. reinhardtii* are well-described [20–22]. The photosystem I of this algae has been shown to contain nine LHCs where LHCA2, 4 and 9 were found to have the greater amount of *red forms*[23] (i.e., chlorophylls with a redshifted light behavior) that lightly enlarge their absorption spectrum.

The potential for using protein-based dyes (e.g., LHCs from *C. reinhardtii*) instead of traditional synthetic light-harvesters is enabled by their feasibility to be produced by recombinant expression in model microorganisms[14]. Research in this area is, however, still at its infancy and only a few studies have attempted to explore this potential[9,24]. An advantage of recombinant LHCs is the possibility to develop protein engineering routes to improve their activity and facilitate their controlled deposition or immobilization within DSSCs. Thus far, specific tags have been only added to LHCs to facilitate their downstream processing[13,14,25], however, the potential of using tags to favor specific immobilization/orientation remains largely unexplored.

As mentioned before, in order to drive a photovoltaic behavior in DSSCs, an electrode is also required to separate the light-induced charge from LHCs. Traditionally, DSSCs have made use of 2D semiconductor materials as electrodes mainly because they have been widely studied at a fundamental level. Some of the semiconductor materials that have been successfully assembled into DSSCs include quantum dots[25], carbon nanotubes[26], and nanoparticles such as $ZnO$[27] and $TiO2$[8].

Recently, graphene has been used as electrode due to its unique properties (e.g., high electrical conductivity and large carbon lattice structure), which can be even further tuned according to its specific application[28]. Even though pristine graphene is not a semiconductor, a controlled mechanism such as surface oxidation might be used to open its electronic band-gap, which in turn allows a further surface charge separation[29–31]. Thus far, no reports are available on the potential of graphene as a substrate for the immobilization of light-harvesting proteins. This work is therefore aimed at rationally designing hybrid photo-electrochemical devices using graphene-based electrodes as a platform for the immobilization of recombinantly-expressed tagged *C. reinhardtii* LHCA proteins for solar energy capture and photocurrent generation.

## RESULTS AND DISCUSSION

### A rationally engineered platform for the production of *C. reinhardtii* light-harvesting proteins in *E. coli* was assembled

A current challenge for the application of light-harvesting complexes in DSSCs is the development of strategies for low-cost production of such proteins at large scale, which might eventually facilitate their study and field implementation. In this regard, microbial biofactories have been proposed as potential platforms for the production of light-harvesting complexes[13,24,25,32]; however, research and development in this area is still in its infancy. Thus, the majority of reported studies have used light-harvesting complexes from bacteria and higher plants, and little is known about the potential of light-harvesting complexes from other photosynthetic microorganisms[14–16]. In this work, a biofactory for the heterologous production of *C. reinhardtii* light-harvesting proteins in *E. coli* has been rationally designed and constructed. Initially, two expression cassettes (Cassette1 and Cassette2) were designed based on two configurations as illustrated in Fig. 1a. Both cassettes were placed on expression vectors pET6xHN-C and pET6xHN-N as described above. The topology of the resulting plasmids is shown in Fig. 1b. These plasmids (pET6xHN-C::Cassette1 and pET6xHN-N::Cassette2) were used to construct the strains *E. coli*_pETC-*LHCA4* and *E. coli*_pETN-*LHCA4*, respectively. After inducing both cultures with IPTG, the purified proteins were evaluated by SDS-PAGE. Fig. 1c confirms the presence of proteins C-LHCA4 (from *E. coli*_pETC-*LHCA4*) and N-LHCA4 (from *E. coli*_pETC-*LHCA4*) with predicted weights of 32.7 kDa and 33.7 kDa, respectively. This agrees well with the reported molecular weight for the proteins[33].

For most of the reported DSSCs that incorporate photosynthetic reaction centers as active transduction components, the light-harvesting complexes or photosystems have been directly isolated from organisms[10], and only a few reports have successfully expressed such proteins in microorganisms[14]. In this work, Leu and Lys tags have been added for the first time to light-harvesting proteins in order to promote both the desired orientation and a controlled immobilization, as explained below. The *plug-and-play* design of the constructed platform may allow the highly efficient expression of additional native or synthetic light-harvesting proteins. This opens opportunities to further study the energy capturing and transduction performance of novel photosynthetic proteins, which could be either engineered from existing or recently isolated microorganisms.

**EEG-based electrodes as suitable platforms for light-harvesting proteins immobilization**

As discussed above, with the purpose of assembling photo-electrochemical devices, the engineered proteins (photosynthetic bio-inspired dyes) must be first immobilized on electrodes (i.e., photoanodes) to enhance light-induced charge separation[34–36]. Within the 2D materials that could be used as electrodes, graphene substrates have unique properties such as high electrical conductivity and carbon lattice structure that can be tuned according to the application[37–39]. Surprisingly, there are no reports regarding the potential of graphene as a substrate for the immobilization of light-harvesting proteins. In this work, a graphene-based electrode (Fig. 2a) was used as a platform for the immobilization of *C. reinhardtii* C-LHCA4 and N-LHCA4 proteins.

Oxygen plasma treatment was applied to generate a reactive surface on graphene, which enabled the formation of carboxyl and epoxy functional groups (C-O-C) on the surface of graphene thin films (here referred to as EEG-based electrode)[29–31]. This method introduces lattice defects with electronegative groups that break the lattice symmetry leading to a p-type doping behavior of the surface where conductive regions ($sp_2$ hybridization) and non-zero bandgap regions ($sp_3$ hybridization) coexist on the same lattice[31]. The level of surface alteration on graphene using oxygen plasma with different exposure times was characterized by Raman spectroscopy (Fig. 2b). Surface alterations are determined by the intensity ratios of graphene characteristic peaks D and G (I(D)/I(G)). D and G peaks are related to the lattice vibrational modes: $1350 cm^{-1}$ for D mode (in-plane vibrations of $sp_2$ carbon atoms) and $1580 cm^{-1}$ for G mode (in-plane vibrations of $sp_3$ carbon atoms). Oxygen plasma was expected to change carbon hybridization from $sp_2$ to $sp_3$, thereby decreasing the I(D)/I(G) ratio when exposure time increases (Fig. 2b). This result agrees with previously reported observations[29,30].

The surface chemistry of the designed EEG-based electrodes was evaluated using XPS as illustrated in Fig. 2c. Binding energies between 200-1200eV reveal C1s (250-350 eV) and O1s (500-550 eV) core energy levels for all tested exposure times (0, 30, 60, 120 and 180s). When the exposure time exceeds 60s; however, the binding energies of the In $3p_{1/2}$ (710 eV) and In $3p_{3/2}$ (724 eV) photoelectron lines from the ITO begin to have significant input as observed for exposure times of 120s and 180s. This observation was not detected by Raman spectroscopy due to the lower resolution of this technique when compared with XPS. The observed binding energies In $3p_{1/2}$ and In $3p_{3/2}$ determined a change in the surface chemistry of EEG-based electrodes, which suggests a significant lessening of EEG from the electrode surface. Such exposure times (where ITO binding energies are detected) were not considered in this study to assure a maximal level of immobilized proteins.

To elucidate changes in surface chemistry, a high-resolution XPS (HR-XPS) around the C1s for EEG-based electrodes at different exposure times was performed (Supplementary Fig. S2). Data showed to be similar to previous reports on reduced graphene oxide[40] and indicated the presence of O-C=O groups, C-O-C groups, and the already present $sp_2$-(C=C) and $sp_3$-(C-C) hybridized carbon atoms. The C1s HR-XPS for the EEG-based electrode with 60s of time exposure is shown in Fig. 2d. The number of functional groups observed in C1s spectra was found to be similar for all the oxygen plasma treatments (Supplementary Fig. S2). Due to the absence of detrimental changes in the surface chemistry, we selected 60s as the optimal plasma exposure time to prepare the EEG-based electrodes. The ease and versatility to control the surface chemistry make the EEG-based electrodes a suitable platform for further protein immobilization.

**Functional EEG-based PEC devices based on *C. reinhardtii* light-harvesting proteins were assembled**

The general aim of this work was to assemble photo-electrochemical devices using heterologously-expressed *C. reinhardtii* light-harvesting proteins C-LHCA4 and N-LHCA4 immobilized on EEG-based substrates. Proteins were designed to fulfill the following two conditions: (i) the photo-active region of proteins must be oriented towards the light source when immobilized on the EEG-based electrodes. This condition was selected because light-harvesting proteins have been reported to exhibit an orientation-dependent photocurrent [41]. The proper orientation was achieved by using a hydrophobic 10xLeu tag flanking the proteins, as it offers the possibility for a preferential orientation of the photo-active region when suspended in the aqueous electrolyte prior to immobilization. (ii) Proteins must be able to properly conjugate to EEG-based substrates through covalent bonding. The free amine groups from the 10xLys tag also flanking the proteins lead to the formation of an amide bond with the free carboxyl groups on the surface of EEG-based electrodes. By this means, engineered proteins were expected to be properly immobilized and oriented on EEG-based electrodes as illustrated in Fig. 3a. Fluorescence microscopy analysis of immobilized rhodamine-labeled proteins revealed a local even distribution (i.e., over the micrometer range) upon immobilization on the EEG-based electrodes, (Fig. 3b and Fig. 3c, respectively) thereby suggesting successful immobilization and a local homogeneous surface coverage in the microns range (i.e., an effective photoactive area of 0.068 $cm^2$).

Electrodes with immobilized proteins were subsequently used to assemble PhotoElectroChemical cells (PEC devices) using different electrolyte compositions and photoanode configurations (Fig. 4a and Supplementary Fig. S1). PECs photocurrent measurements revealed a typical anodic behavior in all tested devices (Fig. 4b). The recorded photocurrent through a light-on cycle showed similar behavior to that observed for charging current in an RC circuit. Similarly, the current in a light-off cycle is comparable to the behavior of the discharging current in an RC circuit as shown in Fig. 4b. This photocurrent behavior has been observed in a wide variety of photovoltaic devices[42–44].

At first, PEC devices were assembled, in the absence of LHCA4 proteins by using different electrolytes and, subsequently in the presence of these proteins. For both set of devices, photocurrent measurements in PEC devices assembled in the absence of both protein and PV revealed the presence of a background photocurrent (Fig. 4c), which could be attributed to the basal activity of electrodes. In this regard, it has been recently shown that graphene display photocurrent in isolation[45,46]. Basal photocurrent also varied according to the electrolyte composition within the microampere range (Fig. 4c). Additionally, photocurrent was measured in PEC devices where PV was stacked and compared to devices where PV was in suspension (Fig. 4c). Devices with stacked PV showed higher basal photocurrents as compared to devices with PV in suspension regardless of the electrolyte composition. This might be explained by the coulombic interactions of reduced PV species near the space-charge layer that lead to their oxidation and transport of the extracted electron through the external circuit. This result agrees well with an observed red-shift in the absorption spectrum of the PV molecule when stacked on the EEG-based electrode (Supplementary Fig. S3). Another likely explanation of the high photocurrent observed could be related to the thickness of the surface coating of PV, which in turn might be due to the used dip coating deposition technique, which might ultimately endow a more pronounced semiconductor-like behavior to the EEG-based electrode [47]. Consequently, the observed photocurrent for devices where PV was in suspension was lower when compared to devices with stacked PV. This can be explained by the fact that a lower amount of suspended PV resulted in a reduced number of charge separation reactions in the vicinity of the photoanode.

In order to determine the role of engineered proteins on charge separation events, photocurrent was measured for PEC devices in the presence of protein (Fig. 4d). These devices were assembled using 0.1M KCl and 0.1M PV as PV-supplemented electrolyte which was set up as the fixed condition in the second set of DSSCs devices due to both the previously reported interaction between light-harvesting proteins and viologen molecules [25] and the change in transmittance when using $I_2$ and PV in the same electrolyte solution. As light-harvesting proteins require pigments to enable charge transfer upon light exposure[15], pigments from *C. reinhardtii* were isolated and subsequently used for protein reconstitution. Chlorophyll a and b and carotenoids were extracted from algae pellets as described in the experimental section. The reconstitution of proteins with pigments was carried out after protein immobilization to avoid photobleaching of pigments[48]. Even though the electrolyte composition was fixed in these protein-based devices, pigments were placed on PECs at three different points during the preparation of the devices considering that such a dependence has not been previously reported (Fig. 4d): 1) along with proteins in solution before immobilization, 2) diluted in the electrolyte after immobilization, and 3) added to the PEC by both methods. Our data showed an enhanced photocurrent in protein-based PECs containing pigments as compared to protein-based PECs alone (Fig. 4d). This observation suggests a light-induced charge transfer from the proteins to the photoanode (Supplementary Fig. S4). This agrees well with recent experimental evidence that indicates that light-harvesting proteins play a role in charge transfer that is different from that observed in photosynthesis[25]. DSSCs based on photosynthetic proteins have been observed to generate a photocurrent because they are assembled with photosystems I or II where charge transfer has been described to be driven by the reaction center[49]. In this study, the charge separation reaction takes place at the interface with the electrode where the energy from extracted pigments (i.e., chlorophylls and carotenoids) has been funneled by the LHC and subsequently separated at the surface of the graphene thin-film. This process is driven by the proteins in the presence of an electric potential gradient (i.e. the space charge layer on the EEG surface). This interaction agrees with the observation that pigments alone fail to generate any photocurrent (Supplementary Fig. S5).

Additional absorption spectroscopy of the assembled photoanodes was carried out to study interaction between the LHCA4 proteins and the pigments (Supplementary Fig. S3). The absorption spectra in each device was de-convoluted to identify a likely interaction, however, no red-shift was observed. This result agrees well with the small photoactive area which might be responsible for the scarce light-transfer events at the interface thereby leading to no observable interaction. Moreover, the pigments were assembled with the LHCA4 proteins without an unfolding experimental procedure thus leading to a relatively weak interaction. In this regard, additional characterization of low-range interactions is required to understand the charge-transfer mechanism between the pigments throughout the LHCA4 protein.

Photocurrent on protein-based PECs was observed to depend on the assembled protein, which seems to suggest a strong dependence on its orientation[41]. The protein C-LHCA4 was designed to locate the N-terminal of LHCA4 towards the light source while the protein N-LHCA4 was designed to position backward. The photocurrent was found to be higher in devices with immobilized C-LHCA4 regardless of the step at which the pigments were added to the PECs (Fig. 4d). The crystallographic structure and photoactive region of the photosystem I of *C. reinhardtii* have been elucidated very recently[20]. Our results agree with Xiaodong et al. observations according to which light-capturing proceeds from the N-terminal region of the LHCA4 complex. For this reason, the complex is oriented towards the chloroplast stroma where light reaches the photosystem I. Our engineered design with Lysine and Leucine tags flanking the proteins appeared useful to immobilize LHCA4 proteins in an orientation for maximal light capturing and transfer.

Furthermore, photocurrent magnitude varied depending on the step where pigments were added during the PEC assembly (Fig. 4d). The highest photocurrent density (40.30 ± 9.26 $\mu A/cm^2$) was obtained for the PEC where pigments were reconstituted during immobilization of the protein complex. It is important to mention that proteins were not entirely distributed over the thin films and some areas remained uncoated, which leaves room to further improvement in the photocurrent. Additionally, when pigments were added to the electrolyte after immobilization (Fig. 4d), the resulting photocurrent was the lowest.

This result might be likely a consequence of the partial assembly of the pigments, most likely due to the reduced solubility and high surface tension. In addition, incubation of immobilized proteins with pigments might promote interactions with residues involved in the energy transfer[10,50,51]. Finally, the photocurrent in PECs assembled with pigments just before and after immobilization was observed to be between both –Chlls and +Chlls, which suggests that protein functionalization is affected by the time pigments interacted with the protein before measurement.

Incident Photon-to-electron Conversion Efficiency action spectra (IPCE) for the devices C-LHCA4 and N-LHCA4 were recorded using a similar setup (Supplementary Fig. S6). The external quantum efficiency reached values of around 50% and 60%, respectively, for wavelengths between 800-900 nm. As a point to be raised, illumination through the counter-electrode led to a higher IPCE between 400 and 700 nm for the N-LHCA4 device. This swap of the dominant wavelengths depending on the direction of illumination supported anew the effect of the protein orientation on the photocurrent performance. Additionally, an estimated Absorbed Photon-to-electron Conversion Efficiency action spectra (APCE)[52] was observed to display a similar behavior when compared to the IPCE. This observation might be due to the low amount of immobilized proteins on the photoanodes (Supplementary Fig. S3).

**CONCLUSION**

The possibility of manufacturing low-cost graphene-based DSSC with rationally-designed light-harvesting proteins from photosynthetic microorganisms enables the next generation of highly efficient biohybrid photoelectrochemical cells. In this work, Leu and Lys tags have been added for the first time to light-harvesting proteins to promote their preferential orientation during immobilization on graphene electrodes. The modularity of the developed platform allows for rational engineering of *plug-and-play* PECs. This is critical to rapidly evaluate other natural and engineered photosynthetic proteins and cell configurations. Indeed, additional work on the photoelectrode performance including both external and internal quantum efficiency is required to explore the photovoltaic potential of the DSSCs proposed in this study. This work also opens the possibility to further explore fundamental processes of energy transfer for biological components interfaced with. To capitalize on these new possibilities there is still a need to overcome the challenge of low structural stability and integrity of the protein complexes as well as the yield of dye complexation [12]. The production of *C. reinhardtii* light-harvesting proteins in *E. coli* is proposed as a suitable strategy to enable the low-cost production required for applications at a larger scale. Further work is nevertheless needed to secure a reliable and high-yield downstream processing. Finally, EEG was shown to be not only a suitable charge separator but also an ideal substrate for the immobilization of engineered proteins for biohybrid DSSCs.

**EXPERIMENTAL SECTION**

**Media composition**

*E. coli* cells were grown on LB medium containing 1% (w/v) tryptone, 1% (w/v) sodium chloride and 0.5% (w/v) yeast extract. Ampicillin and IPTG were added to the media at final concentrations of 100 µg/mL and 1 mM, respectively. The microalgae *C. reinhardtii* was routinely grown on TAP medium (1% (w/v) 5X Beijerinck's solution, 0.83% (w/v) Phosphate solution, 1% (w/v) Tris-acetate solution and 0.1% (w/v) of Trace Elements solution) as described by Elizabeth Harris in 2009 [53]. Solid media were prepared with agar-agar at concentrations of 1.5% (w/v).

**Strains and plasmids construction**

All strains used in this study are listed in Table 1. Two expression cassettes containing the *LHCA4* gene form *C. reinhardtii* were designed, codon-optimized for expression in *E. coli* and subsequently synthesized (Shanghai ShineGene Molecular Biotech, Beijing, China).

DNA sequences for each designed cassette are shown in Supplementary Table S1. Cassette1 was designed to contain 5´-Leu and 3´-Lys tags flanking the *LHCA4* gene. In contrast, Cassette2 was designed to contain a 5´-Lys and 3´-Leu tags flanking the same gene. Cassette1 was subcloned into the expression vector pET6xHN-C (Clontech Laboratories, California, USA) using the restriction enzymes *Pst*I and *Hind*III, and the T4 ligase following standard protocols[54] to obtain the plasmid pET6xHN-C::Cassette1. Similarly, Cassette2 was subcloned into the expression vector pET6xHN-N using the restriction enzymes *Xba*I and *Pst*I to obtain the plasmid pET6xHN-N::Cassette2. Finally, each construct was verified by Sanger sequencing.

**Heterologous expression of recombinant LHCA4 proteins in *E. coli***

Plasmids pET6xHN-C::Cassette1 and pET6xHN-N::Cassette2 were then transformed into *E. coli* BL21(DE3) Gold following the heat shock standard protocol[55]. Positive transformants were selected on solid LB medium supplemented with 100 ug/mL ampicillin. For heterologous expression, the strains *E. coli*_pETC-*LHCA4* and *E. coli*_pETN-*LHCA4* (Table 1) were grown until an $OD_{600nm}$ between 0.5-0.6 was reached. Subsequently, cultures were induced with 1 mM IPTG for 4 hours.

Protein purification was carried out by affinity chromatography using a ProfinityTM IMAC Resin Ni-charged (Bio-Rad) and following standard protocols [56]. Briefly, culture-pellets were obtained by centrifugation at 1400 x g at 4°C for 15 minutes and fully re-suspended in Sonication buffer (0.69% (w/v) $NaH_2PO_4·H_2O$, 1.75% (w/v) NaCl, 0.01% (v/v) Tween 20, 0.001% (v/v) Triton and Lysozyme 1mg/mL, pH=8.0). Samples were lysed using a VibraCell sonicator (Sonics & Materials, USA) at an amplitude of 37% for 13.33 s through 40 cycles of 20s of sonication and 40s of pause. For each induced strain, proteins C-LHCA4 (from *E. coli*_pETC-*LHCA4*) and N-LHCA4 (from *E. coli*_pETC-*LHCA4*) were collected from the supernatants and purified through an equilibrated Ni-charged resin after 30min incubation at 4°C. Fractions of 7 mL were collected for washing and elution steps for subsequent SDS-PAGE analysis following the standard protocols [57].

**Graphene electrode preparation**

Graphene aqueous dispersions were prepared according to the standardized method of electrochemical exfoliation [58]. Thin films of graphene were deposited by spray-coating on glass substrates with an Indium Tin Oxide layer (ITO, Ossila Ltd.) at 150°C (0.09 μg/mm2). A copper wire was placed on a 3mm x 15mm edge region of the transparent electrode with silver paint for a mean illumination area of 0.068 ± 0.015 $cm^2$.

Due to the absence of accessible functional groups on the surface of pristine graphene, further biological functionalization by covalent conjugation represents a major challenge [31]. To overcome this issue, the deposited thin film of graphene was activated by oxygen plasma treatment to induce surface carboxyl and epoxy groups [29,30] This is in contrast to the traditional Hummer's method for the preparation of graphene oxide, which is the common approach to facilitate functionalization [59]. Oxygen plasma treatment was performed with a power and oxygen purity of 22W and 99.7%, respectively using exposure times in the range of 0-180s. The effect of the treatment on the Electrochemical Exfoliated Graphene (EEG) lattice was evaluated via Raman using an excitation wavelength of 785 nm and X-ray photoelectron spectroscopy (XPS).

**LHCA4 immobilization on the graphene surface**

Protein immobilization was carried out in an aqueous solution with the aid of 1-ethyl-3-(3-dimethylaminopropyl)-carbodiimide (EDC) and N-hydroxysuccinimide (NHS) to form amide bonds between the carboxyl groups on the surface of plasma-treated graphene and the pendant amine groups on the LHCA4 protein. To conduct the conjugation, 0.04% (w/v) of protein dissolved in elution buffer was mixed under magnetic stirring at 100 rpm with 1.36% (w/v) of EDC and 1.36% (w/v) of NHS. The plasma-treated EEG thin-film was then placed on the mobile arm of a lab-made dip-coater leading to a protein immobilization efficiency similar to that obtained via Langmuir Blodgett film deposition method [60]. Immersions of the thin-film into each aqueous sample were performed at 0.5 mm/s at 0-16°C for 24h (~1400 cycles) with the protein solution under magnetic stirring at 100 rpm. Reactants excess was removed by additional dipping in deionized water (dH$_2$O) and thin-films were dried out in a low-vacuum chamber at 718 mbar. Samples were stored at 4°C until further use. Similarly, Rhodamine labeled proteins were immobilized for subsequent evaluation of the protein coverage under a ZEISS fluorescence microscope at a wavelength around 532 nm. The labeling was carried out by additional mixing of dissolved protein, EDC and NHS as previously mentioned with 1.36% (w/v) of Rhodamine. Samples were incubated at 37°C for 15 minutes and subsequently let to react under magnetic stirring at 100 rpm for 24h before conducting the conjugation.

**Viologen immobilization on the graphene surface**

Viologens have been explored as alternative dyes in photoelectrochromic cells due to their ability as redox mediators to lead electronic charge transfer in the presence of prokaryotic light-harvesting proteins [25]. Here we incorporated the phosphonated viologen (PV) N-(Diethylphosphono-2-ethyl)-4,4'-bipyridinium bromide as both a redox mediator and a light absorber. The PV was synthesized according to previous reports [61]. Briefly, 1.5 g of 4,4'-bipyridine (Sigma-Aldrich) was diluted in 5.25 mL of 97% diethyl 2-bromoethylphosphonate (AllScience Colombia) and 10mL of Dimethylformamide (DMF) and subsequently maintained for 24 h at 50°C under continuous agitation. The synthesis product was cooled down, washed with 20 mL of diethyl-ether and dried out in a low-vacuum rotary evaporator for 36h. The resulting product was collected and finally stored at room temperature until further use.

EEG thin-films deposited on ITO were immersed in 4%(w/v) of PV dispersed in dH$_2$O. They were let to react for 18h at 50°C under magnetic stirring. Reactants excess was removed by additional dipping in dH$_2$O and the obtained thin-films were dried out in a low-vacuum chamber at 718 mbar. Samples were stored at 4°C until further use.

**Assembly of PEC devices**

The EEG-based electrodes with immobilized proteins C-LHCA4 and N-LHCA4 were assembled with different electrolytes composition. The aqueous electrolyte was supplemented with 0.1M KCl, 0.1M I$_2$ and 0.1M PV and the composition varied according to the presence or absence of each component to identify photocurrent resulting from both proteins and electrodes. The assembly was carried out by placing together the EEG-based electrode as the photoanode and an ITO-substrate as the cathode and electrically connected by the electrolytic interface in-between. Additionally, PV was placed onto the EEG-based electrodes by spontaneously sp$_2$ stacking with simultaneous immobilization of the proteins C-LHCA4 and N-LHCA4 to allow a charge-transfer interaction.

Light-harvesting proteins exhibit light-induced reactions only when pigments are appropriately distributed along with the protein structure [15]. Thus, proteins were driven through a reconstitution step with pigments after immobilization as explained in more detail below. Initially, chlorophylls a and b, and carotenoids were extracted from *C. reinhardtii* pellets obtained by centrifugation of cell cultures at 4°C and 1400 x g for 5 minutes. The obtained pellets were freeze-dried in nitrogen, macerated and subsequently resuspended in 96% ethanol by vigorous vortexing. Soluble pigments were then isolated from the remaining biomass by centrifugation at 4°C and 16300 x g for 2 minutes. Pigments concentration was determined using an UV-vis spectrophotometer at 664 nm for chlorophyll a (Chla), 649 nm for Chlb and 470 nm for carotenoids as described by Sumanta et al. [62]. Samples were stored at -20°C for further use.

The EEG-based electrodes (photoanode) were assembled with the expressed proteins and PV under different photoanode configurations (Supplementary Fig. S1). Briefly, PECs were assembled using the following five configurations: a) EEG-based electrodes and ITO substrate closed by the electrolyte as experimental control, b) stacked PV onto photoanode also closed by the electrolyte, c) diluted PV on electrolyte, and finally immobilized proteins (d) and immobilized proteins with dispersed PV in the electrolyte (e).

**Photocurrent measurements**

Photocurrent measurements were performed using a Keithley source meter 2450 based on the photocurrent decay method (PCD) at a constant applied voltage of 1mV. Copper wire contacts from the photoanode and cathode on assembled PECs were connected to the source meter allowing electron transport through an external circuit [63]. Measurements were performed at room temperature for light-on and light-off cycles using a Xenon lamp MAX-303 as the light source at a constant power of 99.5 mW/$cm^2$. The photocurrents of the representative replicate were used for subsequent analysis. Current in dark condition was measured for 5000ms, to equilibrate the flux of charge carriers, before the first light-on and light-off cycle. IPCE analysis was performed with a commercial apparatus (Arkeo – Cicci Research s.r.l.) using a monochromatic light source (300–1100 nm) based on monochromator and a 150W Xenon lamp with a thermal controlled stage.

**ACKNOWLEDGMENTS**

We thank Universidad de Los Andes (Faculty of Sciences & Department of Chemical Engineering) for financial support, and the researchers at Grupo de Diseño de Productos y Procesos (GDPP) and Nanomaterials Laboratory for their constant contributions, correct observations, and great knowledge. We also thank Javier Cifuentes for his preliminary work.

**AUTHOR CONTRIBUTIONS**

Formal analysis & Data curation: MOT, YH, MFN, AC and FM. Funding acquisition: YH, EJP and AFGB. Investigation and Methodology: MOT, AC, FM and MFN. Project administration: JCC, YH and AFGB. Supervision: JCC, MFN, YH and AFGB. Writing ± original draft: MOT and MFN. Writing ± review & editing: MOT, MFN, AC, FM, JCC, EJP, YH and AFGB.

**COMPETING INTERESTS**

Declarations of interest: none. This research was conducted in the absence of any commercial or financial or other relationships that could be construed as a potential conflict of interest. All authors have approved the final article.

**DATA AVAILABILITY**

All data generated or analyzed during this study are included in this published article (and its Supplementary Information files).

**REFERENCES**

1. Tan, S. C. *Photosynthetic Protein-based Photovoltaics*. (CRC Press, 2018).
2. Dresselhaus, M. S. & Thomas, I. L. Alternative energy technologies. *Nature* **414**, 332 (2001).
3. Green, M. A., Emery, K., Hishikawa, Y., Warta, W. & Dunlop, E. D. Solar cell efficiency tables (Version 45). *Prog. Photovolt. Res. Appl.* **23**, 1–9 (2015).
4. de Wild-Scholten, M. M. Energy payback time and carbon footprint of commercial photovoltaic systems. *Sol. Energy Mater. Sol. Cells* **119**, 296–305 (2013).

5. Kommalapati, R., Kadiyala, A., Shahriar, M. & Huque, Z. Review of the life cycle greenhouse gas emissions from different photovoltaic and concentrating solar power electricity generation systems. *Energies* **10**, 350 (2017).

6. Kakiage, K. *et al.* Highly-efficient dye-sensitized solar cells with collaborative sensitization by silyl-anchor and carboxy-anchor dyes. *Chem. Commun.* **51**, 15894–15897 (2015).

7. Andualem, A. & Demiss, S. Review on Dye-Sensitized Solar Cells (DSSCs). *Edelweiss Appli Sci Tech* **2**, 145–150 (2018).

8. Hug, H., Bader, M., Mair, P. & Glatzel, T. Biophotovoltaics: natural pigments in dye-sensitized solar cells. *Appl. Energy* **115**, 216–225 (2014).

9. Kim, Y., Shin, S. A., Lee, J., Yang, K. D. & Nam, K. T. Hybrid system of semiconductor and photosynthetic protein. *Nanotechnology* **25**, 342001 (2014).

10. Yu, D., Zhu, G., Liu, S., Ge, B. & Huang, F. Photocurrent activity of light-harvesting complex II isolated from spinach and its pigments in dye-sensitized TiO2 solar cell. *Int. J. Hydrog. Energy* **38**, 16740–16748 (2013).

11. Yang, Y. *et al.* Effect of the LHCII pigment–protein complex aggregation on photovoltaic properties of sensitized TiO 2 solar cells. *Phys. Chem. Chem. Phys.* **16**, 20856–20865 (2014).

12. Lämmermann, N. *et al.* Extremely robust photocurrent generation of titanium dioxide photoanodes bio-sensitized with recombinant microalgal light-harvesting proteins. *Sci. Rep.* **9**, 2109 (2019).

13. Mozzo, M. *et al.* Functional analysis of phoyosystem I light-harvesting complexes (Lhca) gene products of Chlamydomonas reinhardtii. *Biochim. Biophys. Acta BBA - Bioenerg.* **1797**, 212–221 (2010).

14. Werwie, M., Xu, X., Haase, M., Basché, T. & Paulsen, H. Bio serves nano: biological light-harvesting complex as energy donor for semiconductor quantum dots. *Langmuir* **28**, 5810–5818 (2012).

15. Strümpfer, J., Sener, M. & Schulten, K. How quantum coherence assists photosynthetic light-harvesting. *J. Phys. Chem. Lett.* **3**, 536–542 (2012).

16. Yu, D. *et al.* Enhanced photocurrent production by bio-dyes of photosynthetic macromolecules on designed TiO 2 film. *Sci. Rep.* **5**, 9375 (2015).

17. Arnon, D. I., McSwain, B. D., Tsujimoto, H. Y. & Wada, K. Photochemical activity and components of membrane preparations from blue-green algae. I. Coexistence of two photosystems in relation to chlorophyll a and removal of phycocyanin. *Biochim. Biophys. Acta BBA-Bioenerg.* **357**, 231–245 (1974).

18. Kargul, J., Nield, J. & Barber, J. Three-dimensional reconstruction of a light-harvesting complex I-photosystem I (LHCI-PSI) supercomplex from the green alga Chlamydomonas reinhardtii Insights into light harvesting for PSI. *J. Biol. Chem.* **278**, 16135–16141 (2003).

19. Gantt, E., Grabowski, B. & Cunningham, F. X. Antenna systems of red algae: phycobilisomes with photosystem ll and chlorophyll complexes with photosystem I. in *Light-harvesting antennas in photosynthesis* 307–322 (Springer, 2003).

20. Su, X. *et al.* Antenna arrangement and energy transfer pathways of a green algal photosystem-I–LHCI supercomplex. *Nat. Plants* **5**, 273 (2019).

21. Tokutsu, R., Teramoto, H., Takahashi, Y., Ono, T. & Minagawa, J. The light-harvesting complex of photosystem I in Chlamydomonas reinhardtii: protein composition, gene structures and phylogenic implications. *Plant Cell Physiol.* **45**, 138–145 (2004).

22. Drop, B. *et al.* Photosystem I of Chlamydomonas reinhardtii contains nine light-harvesting complexes (Lhca) located on one side of the core. *J. Biol. Chem.* **286**, 44878–44887 (2011).

23. Le Quiniou, C. *et al.* PSI–LHCI of Chlamydomonas reinhardtii: Increasing the absorption cross section without losing efficiency. *Biochim. Biophys. Acta BBA - Bioenerg.* **1847**, 458–467 (2015).

24. Liu, J., Lauterbach, R., Paulsen, H. & Knoll, W. Immobilization of light-harvesting chlorophyll a/b complex (LHCIIb) studied by surface plasmon field-enhanced fluorescence spectroscopy. *Langmuir* **24**, 9661–9667 (2008).

25. Werwie, M. *et al.* Light-harvesting chlorophyll protein (LHCII) drives electron transfer in semiconductor nanocrystals. *Biochim. Biophys. Acta BBA-Bioenerg.* **1859**, 174–181 (2018).

26. Kaniber, S. M., Simmel, F. C., Holleitner, A. W. & Carmeli, I. The optoelectronic properties of a photosystem I–carbon nanotube hybrid system. *Nanotechnology* **20**, 345701 (2009).

27. Verma, S., Gupta, A., Sainis, J. K. & Ghosh, H. N. Employing a photosynthetic antenna complex to interfacial electron transfer on ZnO quantum dot. *J. Phys. Chem. Lett.* **2**, 858–862 (2011).

28. Xiang, Q., Cheng, B. & Yu, J. Graphene-based photocatalysts for solar-fuel generation. *Angew. Chem. Int. Ed.* **54**, 11350–11366 (2015).

29. Gokus, T. *et al.* Making Graphene Luminescent by Oxygen Plasma Treatment. *ACS Nano* **3**, 3963–3968 (2009).

30. Shin, Y. J. *et al.* Surface-Energy Engineering of Graphene. *Langmuir* **26**, 3798–3802 (2010).

31. Dey, A., Chroneos, A., Braithwaite, N. S. J., Gandhiraman, R. P. & Krishnamurthy, S. Plasma engineering of graphene. *Appl. Phys. Rev.* **3**, 021301 (2016).

32. Paulsen, H., Rümler, U. & Rüdiger, W. Reconstitution of pigment-containing complexes from light-harvesting chlorophyll a/b-binding protein overexpressed inEscherichia coli. *Planta* **181**, 204–211 (1990).

33. Consortium, U. UniProt: a worldwide hub of protein knowledge. *Nucleic Acids Res.* **47**, D506–D515 (2018).

34. Calvin, M. Quantum conversion in photosynthesis. *J. Theor. Biol.* **1**, 258–287 (1961).

35. Grätzel, M. Photoelectrochemical cells. *nature* **414**, 338 (2001).

36. Van de Krol, R. Principles of photoelectrochemical cells. in *Photoelectrochemical hydrogen production* 13–67 (Springer, 2012).

37. De Sanctis, A., Mehew, J., Craciun, M. & Russo, S. Graphene-based light sensing: fabrication, characterisation, physical properties and performance. *Materials* **11**, 1762 (2018).

38. Wei, D. *et al.* Synthesis of N-doped graphene by chemical vapor deposition and its electrical properties. *Nano Lett.* **9**, 1752–1758 (2009).

39. Yan, Z. *et al.* Controlled modulation of electronic properties of graphene by self-assembled monolayers on SiO2 substrates. *ACS Nano* **5**, 1535–1540 (2011).

40. Dreyer, D. R., Park, S., Bielawski, C. W. & Ruoff, R. S. The chemistry of graphene oxide. *Chem. Soc. Rev.* **39**, 228–240 (2010).

41. Kondo, M. *et al.* Photocurrent and electronic activities of oriented-His-tagged photosynthetic light-harvesting/reaction center core complexes assembled onto a gold electrode. *Biomacromolecules* **13**, 432–438 (2012).

42. Danine, A., Manceriu, L., Fargues, A. & Rougier, A. Eco-friendly redox mediator gelatin-electrolyte for simplified TiO2-viologen based electrochromic devices. *Electrochimica Acta* **258**, 200–207 (2017).

43. Nagata, M. *et al.* Immobilization and photocurrent activity of a light-harvesting antenna complex II, LHCII, isolated from a plant on electrodes. *ACS Macro Lett.* **1**, 296–299 (2012).

44. Gadgil, B., Damlin, P., Heinonen, M. & Kvarnström, C. A facile one step electrostatically driven electrocodeposition of polyviologen–reduced graphene oxide nanocomposite films for enhanced electrochromic performance. *Carbon* **89**, 53–62 (2015).

45. Chang-Jian, S.-K., Ho, J.-R., Cheng, J.-W. J. & Hsieh, Y.-P. Characterizations of photoconductivity of graphene oxide thin films. *AIP Adv.* **2**, 022104 (2012).

46. LeBlanc, G., Winter, K. M., Crosby, W. B., Jennings, G. K. & Cliffel, D. E. Integration of photosystem I with graphene oxide for photocurrent enhancement. *Adv. Energy Mater.* **4**, 1301953 (2014).

47. Weiss, T. P. *et al.* Bulk and surface recombination properties in thin film semiconductors with different surface treatments from time-resolved photoluminescence measurements. *Sci. Rep.* **9**, 1–13 (2019).

48. Horn, R. & Paulsen, H. Folding in vitro of light-harvesting chlorophyll a/b protein is coupled with pigment binding. *J. Mol. Biol.* **318**, 547–556 (2002).

49. Lodish, U. H. *et al. Molecular Cell Biology*. (W. H. Freeman, 2008).

50. Hoober, J. K., Eggink, L. L. & Chen, M. Chlorophylls, ligands and assembly of light-harvesting complexes in chloroplasts. *Photosynth. Res.* **94**, 387–400 (2007).

51. Balevičius, V. *et al.* Fine control of chlorophyll-carotenoid interactions defines the functionality of light-harvesting proteins in plants. *Sci. Rep.* **7**, 13956 (2017).

52. Margulis, G. Y., Hardin, B. E., Ding, I.-K., Hoke, E. T. & McGehee, M. D. Parasitic Absorption and Internal Quantum Efficiency Measurements of Solid-State Dye Sensitized Solar Cells. *Adv. Energy Mater.* **3**, 959–966 (2013).

53. Harris, E. H., Stern, D. B. & Witman, G. B. *The chlamydomonas sourcebook*. vol. 1 (Elsevier San Diego, CA, 2009).

54. Lundblad, R. L. & Macdonald, F. *Handbook of biochemistry and molecular biology*. (CRC Press, 2018).

55. Froger, A. & Hall, J. E. Transformation of plasmid DNA into E. coli using the heat shock method. *JoVE J. Vis. Exp.* e253 (2007).

56. Lindner, P. *et al.* Purification of native proteins from the cytoplasm and periplasm of Escherichia coli using IMAC and histidine tails: a comparison of proteins and protocols. *Methods* **4**, 41–56 (1992).

57. Walker, J. M. *The protein protocols handbook*. vol. 1996 (Springer Science & Business Media, 1996).

58. Hurtado-Morales, M. *et al.* Efficient fluorescence quenching in electrochemically exfoliated graphene decorated with gold nanoparticles. *Nanotechnology* **27**, 275702 (2016).

59. Hummers Jr, W. S. & Offeman, R. E. Preparation of graphitic oxide. *J. Am. Chem. Soc.* **80**, 1339–1339 (1958).

60. Kamran, M., Delgado, J. D., Friebe, V., Aartsma, T. J. & Frese, R. N. Photosynthetic protein complexes as bio-photovoltaic building blocks retaining a high internal quantum efficiency. *Biomacromolecules* **15**, 2833–2838 (2014).

61. Campus, F., Bonhote, P., Grätzel, M., Heinen, S. & Walder, L. Electrochromic devices based on surface-modified nanocrystalline TiO2 thin-film electrodes. *Sol. Energy Mater. Sol. Cells* **56**, 281–297 (1999).

62. Sumanta, N., Haque, C. I., Nishika, J. & Suprakash, R. Spectrophotometric analysis of chlorophylls and carotenoids from commonly grown fern species by using various extracting solvents. *Res J Chem Sci* **2231**, 606X (2014).

63. Hodes, G. Photoelectrochemical Cell Measurements: Getting the Basics Right. *J. Phys. Chem. Lett.* **3**, 1208–1213 (2012).

64. Boyd, J., Oza, M. N. & Murphy, J. R. Molecular cloning and DNA sequence analysis of a diphtheria tox iron-dependent regulatory element (dtxR) from Corynebacterium diphtheriae. *Proc. Natl. Acad. Sci.* **87**, 5968–5972 (1990).

65. Pan, S. & Malcolm, B. A. Reduced background expression and improved plasmid stability with pET vectors in BL21 (DE3). *Biotechniques* **29**, 1234–1238 (2000).

66. Palmer, B. R. & Marinus, M. G. The dam and dcm strains of Escherichia coli—a review. *Gene* **143**, 1–12 (1994).

67. Grossman, A. R., Lohr, M. & Im, C. S. Chlamydomonas reinhardtii in the landscape of pigments. *Annu Rev Genet* **38**, 119–173 (2004).

# TABLES

**Table 1.** Strains used in this study.

| Strain name | Description | Reference |
| --- | --- | --- |
| *E. coli* DH5 | Reference laboratory strain | 64 |
| *E. coli* BL21(DE3)Gold | Suitable strain for high-level T7 protein expression | 65 |
| *E. coli* dam-/dcm- | Suitable strain for cloning dam and dcm methylation-free plasmids | 66 |
| *E. coli*_pETC-*LHCA4* | *E. coli* BL21(DE3)Gold strain transformed with the plasmid pET6xHN-C::Cassette1. Cassette1 carries the *LHCA4* gene from *C. reinhardtii* flanked by a 30bp sequence encoding for a 5'-Leu-tag and a 30bp sequence encoding for a 3'-Lys-tag. | This study |
| *E. coli*_pETN-*LHCA4* | *E. coli* BL21(DE3)Gold strain transformed with the plasmid pET6xHN-N::Cassette2. Cassette2 carries the *LHCA4* gene from *C. reinhardtii* flanked by a 30bp sequence encoding for a 5'-Lys-tag and a 30bp sequence encoding for a 3'-Leu-tag. | This study |
| *C. reinhardtii* CC503 cw92mt+ | Model strain to study the photosynthetic machinery. | 67 |

**FIGURE CAPTIONS**

**Figure 1.** Engineered and purified tagged proteins C-LHCA4 and N-LHCA4 from *C. reinhardtii* heterologously expressed in *E. coli*. (a) Schematic representation of synthesized coding sequences for tagged proteins C-LHCA4 (i.e. Cassette1) *(left)* and N-LHCA4 (i.e. Cassette2) *(right)* where the core sequence of the gene *LHCA4* (green) is flanked with a 5'-Leu-tag (yellow) and 3'-Lys-tag (red) for Cassette1, and with a 5'-Lys-tag (red) and 3'-Leu-tag (yellow) for Cassette2. (b) Schematic representation of constructed plasmids pET6xHN-C::Cassette1 *(left)* and pET6xHN-N::Cassette2 *(right)* where a 6xHis tag was added on the C-terminal of Cassette1 and on the N-terminal of Cassette2. (c) SDS-PAGE for purified expressed proteins from *E. coli*_pETC-*LHCA4 (left)* and *E. coli*_pETN-*LHCA4 (right)* where the molecular weights for C-LHCA4 (32.7 kDa) and N-LHCA4 (33.7 kDa) are respectively highlighted.

**Figure 2.** Characterization of EEG-based electrodes. (a) Aqueous dispersion of EEG *(left)* used to fabricate the EEG-based electrodes *(right)*. (b) Raman I(D)/I(G) ratio for electrodes treated with 22W oxygen plasma as a function of oxygen plasma exposure time (0, 30, 60, 120 and 180s). (c) Broad-range XPS spectra (200-1200 eV) of electrodes under the sampled exposure times. (d) Binding energy for deconvoluted HR-XPS C1s for the electrode treated with 60s of plasma exposure time.

**Figure 3.** Immobilization of expressed proteins on EEG-based electrodes based on their molecular structure. (a) Schematic representation of the expected orientation of proteins C-LHCA4 and N-LHCA4 (driven by the 10xLeu tag) when covalently bound to –COOH groups on the surface of EEG-based electrodes through the 10xLys tag. Photo-active region in each case is represented by green dots on the molecular structure. (b) Immobilized C-LHCA4 and (c) N-LHCA4 on EEG-based electrodes under fluorescence microscopy.

**Figure 4.** Photocurrent displayed by assembled PEC devices. (a) Schematic representation of an assembled PEC device with a PV-supplemented electrolyte where the photoanode constitutes immobilized LHCA4 proteins on the EEG-based electrode. The photocurrent flowing through Ag contacts is measured in the external circuit. (b) Typical anodic photocurrent performing an RC behavior through light-on and light-off cycles. (c) Photocurrent densities exhibited by PEC devices without immobilized proteins in the photoanode. Measurements are displayed for PECs assembled with three supplemented aqueous electrolytes: water as the control ($H_2O$), KCl and KCl + $I_2$. Additionally, measurements are displayed for different configurations of assembled PV on PEC devices: without PV as the control (no PV), PV in solution (i.e. PV-supplemented electrolyte, labeled as PV in sln) and stacked PV (PV stacked). (d) Photocurrent densities exhibited by PEC devices with immobilized proteins C-LHCA4 (blue) and N-LHCA4 (green) on the photoanode assembled with a PV-supplemented electrolyte (control measurement is shown with the red dotted line). Measurements are displayed for different pigments (Chlls) assembly on PEC devices: (i) without pigments as the control (no Chlls) (ii) before proteins immobilization (-Chlls), (iii) after proteins immobilization (+Chlls) and (iv) before and after proteins immobilization (-/+Chlls). Associated error bars in the histograms were determined for three sets of several independent light-on and light-off measurements, for each PEC configuration.

# FIGURES

# FIGURE 1

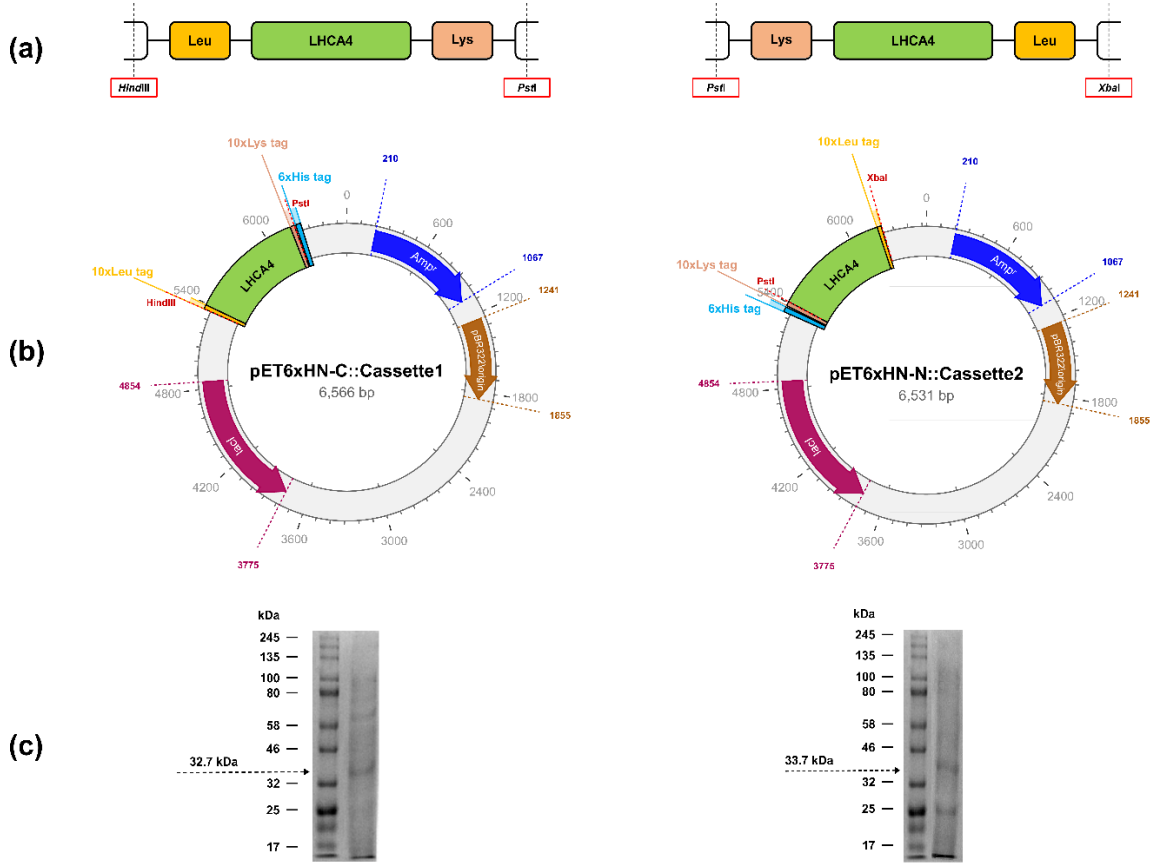

**FIGURE 2**

(a)

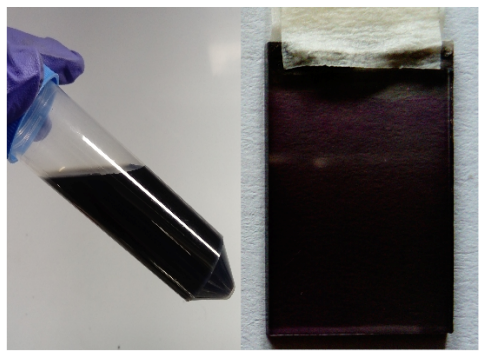

(b)

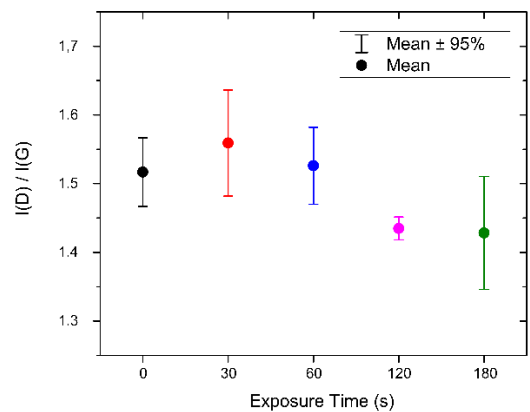

(c)

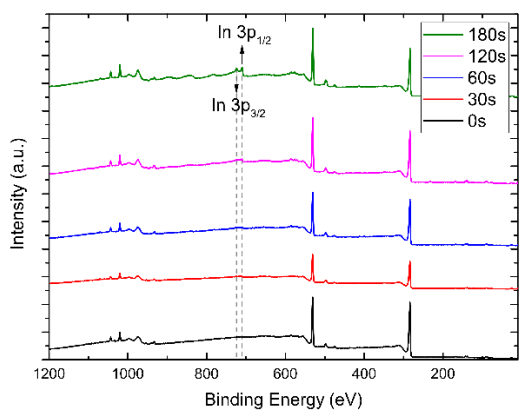

(d)

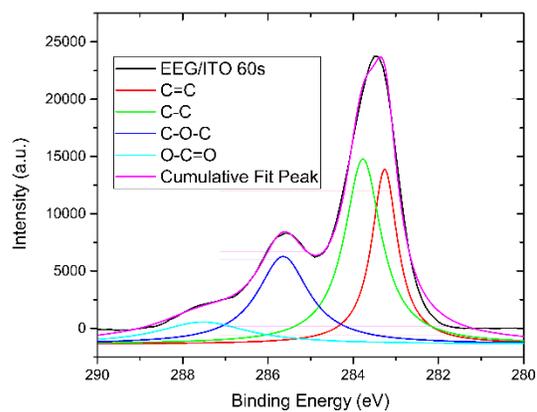

**FIGURE 3**

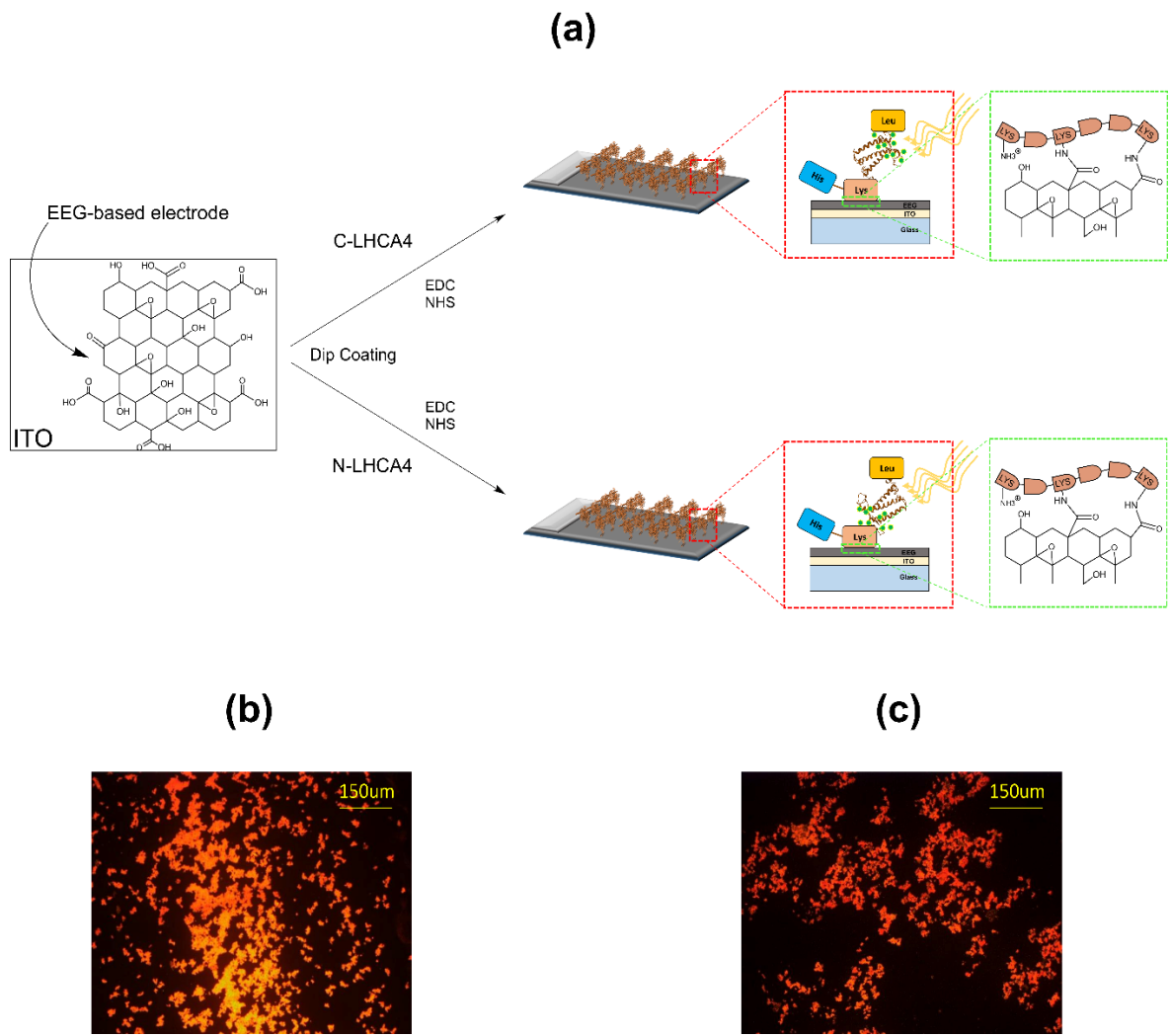

**FIGURE 4**

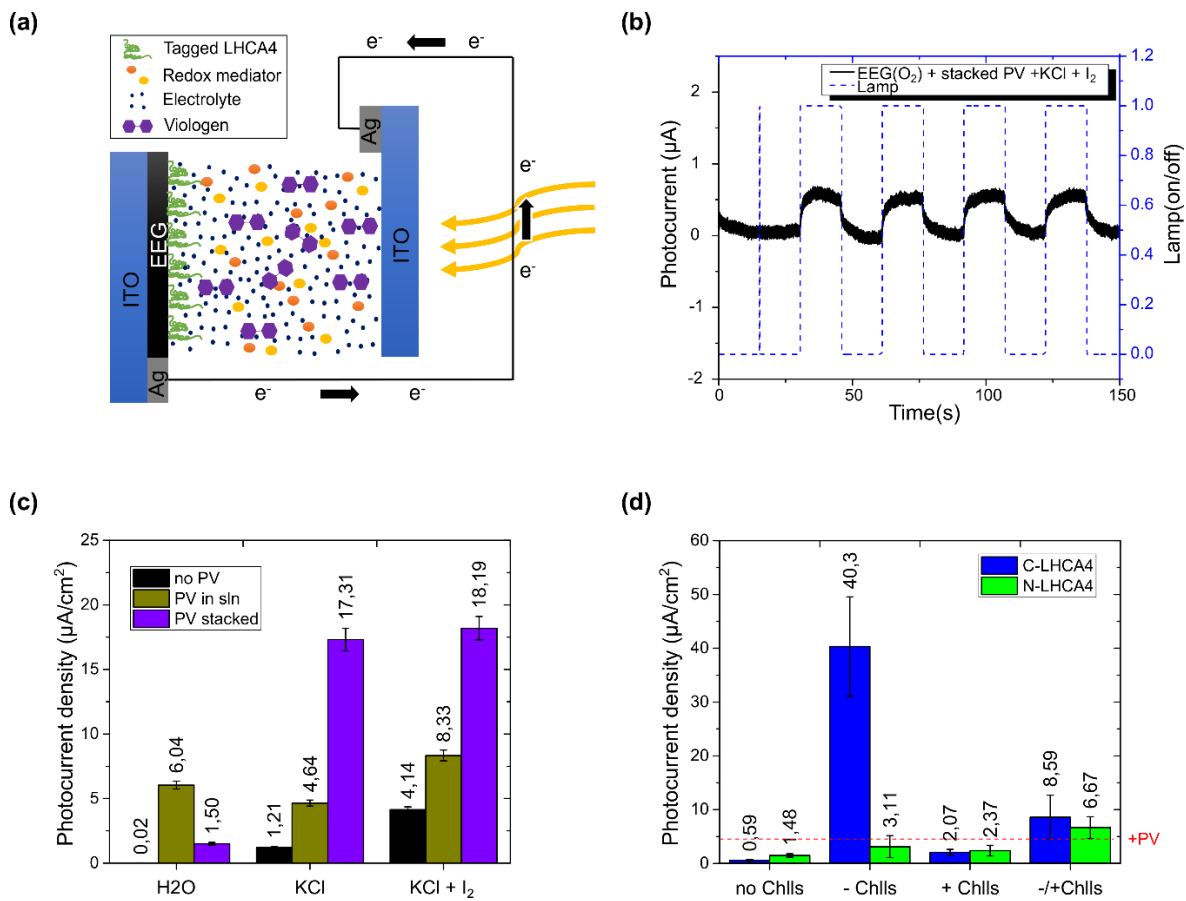